\documentclass[10pt,leqno]{amsart}
\usepackage{graphicx}
\usepackage{indentfirst,csquotes}

\usepackage{amssymb,amsthm,amsmath}
\usepackage{xcolor,paralist,hyperref,titlesec,fancyhdr,etoolbox}

\usepackage{titlesec}
\titleformat{\section}
  {\normalfont\Large\bfseries\centering}
  {\thesection.}
  {0.6em}
  {}
\titlespacing*{\section}{0pt}{2ex}{1ex}

\hypersetup{ colorlinks=true, linkcolor=black, filecolor=black, urlcolor=black }

\DeclareRobustCommand{\VAN}[3]{#2}
\let\VANthebibliography\thebibliography
\def\thebibliography{\DeclareRobustCommand{\VAN}[3]{##3}\VANthebibliography}
\usepackage{hyperref}
\usepackage{amsmath}
\usepackage{layouts}
\usepackage{booktabs}
\usepackage{natbib}
\usepackage{subcaption}
\usepackage{journals}

\newcommand{\FIPrepr}{\mathfrak{R}}

\newcommand\ion[2]{%
  \textup{#1\,{%
    \ifx\@currsize\normalsize\small \else
    \ifx\@currsize\small\footnotesize \else
    \ifx\@currsize\footnotesize\scriptsize \else
    \ifx\@currsize\scriptsize\tiny \else
    \ifx\@currsize\large\normalsize \else
    \ifx\@currsize\Large\large
    \fi\fi\fi\fi\fi\fi
    \textsc{\romannumeral #2}%
  }}%
}

\newcommand\arcsec{\hbox{$^{\prime\prime}$}}

\usepackage[normalem]{ulem} 
\usepackage{xcolor}
\usepackage{xparse}
\usepackage{cancel}         
\usepackage{lineno}

\usepackage{pdftexcmds}  
\usepackage{xstring}     
\usepackage{floatrow}    
\usepackage{caption}
\usepackage{trackchanges-slimane}
\showchangesfalse            

\newcommand{\release}{release~5.0}
\newcommand{\dt}[1]{%
    \ifcase#1\relax
        \or 
        20th 5:08
        \or 
        21st 1:13
        \or 
        22nd 22:53
        \or 
        23rd 21:32
        \or 
        25th 00:56
        \or 
        25th 21:31
        \or 
        26th 8:43
        \else
        Invalid date%
    \fi
    }

\NewDocumentCommand{\parsemydate}{mmmmmm}{%
  \IfNoValueF{#1}{#1}%
  \IfNoValueF{#2}{-#2}%
  \IfNoValueF{#3}{-#3}%
  \IfNoValueF{#4}{ #4}%
  \IfNoValueF{#5}{:#5}%
  \IfNoValueF{#6}{:#6}%
}

\usepackage{pdftexcmds}  
\usepackage{xstring}     

\makeatletter
\providecommand{\@secnumpunct}{.}
\makeatother
\makeatletter

\def\autoFormat{auto}
\def\shortFormat{short}
\def\longFormat{long}

\newcommand{\newacronym}[4][]{ 
  \def\ac@cite{}

  \StrSubstitute{#1}{ }{}[\ac@cite]

  \protected@edef\ac@tempA{#3}%
  \protected@edef\ac@tempB{#4}%
  
  \IfStrEq{\ac@tempA}{\ac@tempB}%
    {\def\ac@same{1}}%
    {\def\ac@same{0}}%

  \expandafter\protected@edef\csname acronym@#2@cite\endcsname{\ac@cite}%
  \expandafter\protected@edef\csname acronym@#2@long\endcsname{#3}%
  \expandafter\protected@edef\csname acronym@#2@short\endcsname{#4}%
  \expandafter\edef\csname acronym@#2@same\endcsname{\ac@same}%
  \expandafter\edef\csname acronym@#2@used\endcsname{0}%
}

\newcommand{\myac}[2][\autoFormat]{
  \StrLen{#2}[\acFullLen]%
  \StrRight{#2}{1}[\lastChar]%
  \StrLeft{#2}{\numexpr\acFullLen-1\relax}[\acMaybePlural]%
  \def\pluralFlag{}%
  \IfStrEq{\lastChar}{s}{\def\pluralFlag{s}}{\def\acMaybePlural{#2}}%
  \StrLeft{\acMaybePlural}{3}[\firstThree]%
  \StrLen{\acMaybePlural}[\maybeLen]%
  \IfStrEq{\firstThree}{the}{%
    \StrRight{\acMaybePlural}{\numexpr\maybeLen-3\relax}[\acBase]%
    \def\printThe{the }%
  }{%
    \IfStrEq{\firstThree}{The}{%
      \StrRight{\acMaybePlural}{\numexpr\maybeLen-3\relax}[\acBase]%
      \def\printThe{The }%
    }{%
      \def\acBase{\acMaybePlural}%
      \def\printThe{}%
    }%
  }%
  \StrSubstitute{\acBase}{ }{}[\acBase]
  \edef\acCite{\csname acronym@\acBase @cite\endcsname}%
  \edef\acUsed{\csname acronym@\acBase @used\endcsname}%
  \edef\acSame{\csname acronym@\acBase @same\endcsname}%
  \def\acLong{\csname acronym@\acBase @long\endcsname}%
  \def\acShort{\csname acronym@\acBase @short\endcsname}%
  \IfStrEq{#1}{\shortFormat}{\def\acFormat{1}}{%
  \IfStrEq{#1}{\longFormat}{\def\acFormat{2}}{%
    \def\acFormat{0}
  }}%
  \ifnum\acFormat=1 
  \else\ifnum\acFormat=2
    \ifnum\acSame=1%
      \ifx\acCite\@empty%
        \printThe\acLong\pluralFlag%
      \else%
        \printThe\acLong\pluralFlag\ \citep{\acCite}%
      \fi%
    \else%
      \ifx\acCite\@empty%
        \printThe\acLong\pluralFlag\ (\acShort\pluralFlag)%
      \else%
        \printThe\acLong\pluralFlag\ \citep[\acShort\pluralFlag;][]{\acCite}%
      \fi%
    \fi%
  \else
    \ifnum\acUsed=0
      \expandafter\gdef\csname acronym@\acBase @used\endcsname{1}%
      \ifnum\acSame=1%
        \ifx\acCite\@empty%
          \printThe\acLong\pluralFlag%
        \else%
          \printThe\acLong\pluralFlag\ \citep{\acCite}%
        \fi%
      \else%
        \ifx\acCite\@empty%
          \printThe\acLong\pluralFlag\ (\acShort\pluralFlag)%
        \else%
          \printThe\acLong\pluralFlag\ \citep[\acShort\pluralFlag;][]{\acCite}%
        \fi%
      \fi%
    \else
      \printThe\acShort\pluralFlag%
    \fi%
  \fi\fi%
}%
\makeatother

\newacronym[]{ESA}{European Space Agency}{ESA}
\newacronym[]{NASA}{National Aeronautics and Space Administration}{NASA}
\newacronym[]{JAXA}{Japan Aerospace Exploration Agency}{JAXA}

\usepackage{xspace}
\newacronym[SOLO]{SOLO}{Solar Orbiter}{Solar Orbiter}
\newcommand{\solo}{\myac{SOLO}\xspace}
\newacronym[SOHO]{SOHO}{Solar and Heliospheric Observatory}{SoHO}
\newacronym[SUMER]{SUMER}{Solar Ultraviolet Measurements of Emitted Radiation}{SUMER}
\newacronym[Hinode]{Hinode}{Hinode}{Hinode}
\newacronym[CDS]{CDS}{Coronal Diagnostic Spectrometer}{CDS}
\newacronym[ACE]{ACE}{Advanced Composition Explorer}{ACE}
\newacronym[SWICSUlyssis]{SWICS}{Solar Wind Ion Composition Spectrometer}{SWICS}
\newacronym[EPACT]{EPACT}{Energetic Particles Acceleration, Composition, and Transport}{EPACT}
\newacronym[IMP8]{IMP8}{Interplanetary Monitoring Platform-8}{IMP-8}
\newacronym[GST]{GST}{Goode Solar Telescope}{GST}
\newacronym[NIRIS]{NIRIS}{Near-InfraRed Imaging Spectropolarimeter}{NIRIS}
\newacronym[]{BBSO}{Big Bear Solar Observatory}{BBSO}
\newacronym[DKIST]{DKIST}{ Daniel K. Inouye Solar Telescope }{DKIST}
\newacronym[SolarC]{SOLARC}{Solar-C}{Solar-C}
\newacronym[EST]{EST}{ European Solar Telescope}{EST}
\newacronym[EUVST]{EUVST}{EUV Spectroscopic Telescope}{EUVST}
\newacronym[EUI]{EUI}   {Extreme-Ultraviolet Imager}               {EUI}
\newacronym[EUI]{FSI}   {Full-Sun Imager}                          {FSI}
\newacronym[EUI]{HRI}{High-Resolution Imager}{HRI}
\newacronym[PHI]{PHI}   {Polarimetric and Helioseismic Imager}{PHI}
\newacronym[]{FDT}   {Full-Disk Telescope}{FDT}
\newacronym[]{HRT}   {High-Resolution Telescope}{HRT}
\newacronym[SDO]{SDO}   {Solar Dynamics Observatory}{SDO}
\newacronym[AIA]{AIA}   {Atmospheric Imaging Assembly}{AIA}
\newacronym[HMI]{HMI}   {Helioseismic and Magnetic Imager}{HMI}
\newacronym[EISInstrument]{EIS}{Extreme-Ultraviolet Imaging Spectrometer}{EIS}
\newacronym[SPICE]{SPICE}{Spectral Imaging of the Coronal Environment}{SPICE}
\newacronym[STIX]{STIX}{Spectrometer/Telescope for Imaging X-rays}{STIX}
\newacronym[SWA]{SWA}{Solar Wind Analyser}{SWA}
\newacronym[PSP]{PSP}{Parker Solar Probe}{PSP}

\newacronym[]{RSW}{Remote-Sensing Window}{RSW}
\newacronym[]{SOOP}{Solar Orbiter Observing Plan}{SOOP}
\newacronym[]{STP}{Short-Term Planning}{STP}
\newacronym[]{MTP}{Medium-Term Planning}{MTP}
\newacronym[]{LTP}{Long-Term Planning}{LTP}
\newacronym[]{VSTP}{Very‑Short‑Term Planning}{VSTP}

\newacronym[IRIS]{IRIS}{Interface Region Imaging Spectrograph}{IRIS}

\newacronym[]{FIP}{first ionization potential}{FIP}
\newacronym[]{IFIP}{inverse FIP}{IFIP}

\newacronym[]{AR}{active region}{AR}
\newacronym[]{CH}{coronal hole}{CH}
\newacronym[]{ROI}{region of interest}{ROI}
\newacronym[]{PDFG}{probability distribution function}{PDF}
\newacronym[]{ICR}{interchange reconnection}{ICR}

\newacronym[]{TR}{transition region}{TR}
\newacronym[]{QS}{quiet Sun}{QS}
\newacronym[]{CME}{coronal mass ejection}{CME}
\newacronym[]{CIR}{corotating interaction region}{CIR}
\newacronym[]{SEP}{solar energetic particle}{SEP}
\newacronym[]{GCR}{galactic cosmic rays}{GCR}
\newacronym[]{EUV}{extreme ultraviolet}{EUV}
\newacronym[]{EFR}{emerging flux region}{EFR}
\newacronym[]{UV}{ultraviolet}{UV}
\newacronym[]{FSW}{fast solar wind}{FSW}
\newacronym[]{SSW}{slow solar Wind}{SSW}
\newacronym[]{AU}{astronomical units}{AU}
\newacronym[]{DEM}{differential emission measure}{DEM}
\newacronym[]{TLR}{two-lines ratio}{2LR}
\newacronym[]{PSF}{point spread function}{PSF}
\newacronym[]{FWHM}{full-width at half-maximum }{FWHM}
\newacronym[]{FOV}{field of view}{FOV}
\newacronym[]{LOS}{line-of-sight}{LOS}
\newacronym[]{SAFFRON}{Spectral Analysis, Fitting Framework, Reduction Of Noise}{SAFFRON}
\newacronym[]{SNR}{signal-to-noise ratio}{SNR}
\newacronym[FITS]{FITS}{Flexible Image Transport System}{FITS}
\newacronym[]{L2}{Level 2}{L2}
\newacronym[]{L3}{Level 2}{L3}
\newacronym[]{WCS}{World Coordinate System}{WCS}

\newcommand{\FSIFe}{\mbox{EUI/FSI\textsubscript{174}}\xspace}

\newcommand{\PHIBLOS}{\mbox{PHI/HRT\textsubscript{BLOS}}\xspace}

\newacronym[LCR]{LCR}{Linear Combination Ratio method}{LCR}

\newcommand{\noyear}{false} 

\NewDocumentCommand{\mydateapj}{O{true} >{\SplitArgument{5}{,}}m}{%
  \parsemydateapj[#1]#2%
}

\NewDocumentCommand{\parsemydateapj}{O{true}mmmmmm}{%
  \IfStrEq{#1}{\noyear}{}{%
    \IfNoValueF{#2}{#2 }
  }%
  \IfNoValueF{#3}{\monthname{#3}}%
  \IfNoValueF{#4}{ #4}%
  \IfNoValueF{#5}{ at #5%
    \IfNoValueF{#6}{:#6}%
    \IfNoValueF{#7}{:#7}}%
}

\newcommand{\monthname}[1]{%
    \ifcase#1\relax
        Invalid month%
        \or January%
        \or February%
        \or March%
        \or April%
        \or May%
        \or June%
        \or July%
        \or August%
        \or September%
        \or October%
        \or November%
        \or December%
    \else
        Invalid month #1%
    \fi
}
\usepackage{graphicx}	
\usepackage{amsmath}	

\usepackage{lipsum}

\begin{document}
\title{Sulfur fractionation in coronal plumes as observed by Solar Orbiter/SPICE} 
\author{Slimane Mzerguat}
\address{Université Paris-Saclay, CNRS, Institut d'Astrophysique Spatiale, 91405 Orsay, France}
\email{slimane.mzerguat@universite-paris-saclay.fr}

\author{Miho Janvier}
\address{ESTEC, European Space Agency, Keplerlaan 1, PO Box 299, NL-2200 AG Noordwijk, The Netherlands}

\author{Eric Buchlin}
\address{Université Paris-Saclay, CNRS, Institut d'Astrophysique Spatiale, 91405 Orsay, France}

\author{Deborah Baker}
\address{Mullard Space Science Laboratory, University College London, Holmbury St Mary, Dorking, Surrey RH5 6NT, UK}

\author{Andy S.H. To}
\address{ESTEC, European Space Agency, Keplerlaan 1, PO Box 299, NL-2200 AG Noordwijk, The Netherlands}

\author{David M. Long}
\address{Centre for Astrophysics \& Relativity, School of Physical Sciences, Dublin City University, Dublin D09 V209, Ireland}
\address{Astronomy \& Astrophysics Section, Dublin Institute for Advanced Studies, Dublin D02 XF86, Ireland}

\author{Natalia Zambrana Prado}
\address{Mullard Space Science Laboratory, University College London, Holmbury St Mary, Dorking, Surrey RH5 6NT, UK}

\date{\today}
{\today}
\email{slimane.mzerguat@u-psud.fr}
\maketitle

\begin{abstract}

Coronal plumes are bright, narrow structures rooted in coronal holes that contribute to the solar wind. Their composition, particularly elemental fractionation as a function of first ionization potential (FIP), provides diagnostics of plasma properties and magnetic connectivity. 
Earlier plume studies of fractionation using low-FIP elements reached conflicting conclusions. Intermediate-FIP elements may provide additional diagnostic insight, since their fractionation is thought to involve processes beyond those affecting low-FIP species.
We investigate sulfur (intermediate-FIP element) in plumes to assess the presence of fractionation, its evolution, and its relation to wave activity.
We analyzed Solar Orbiter observations of two plumes in an equatorial coronal hole during March--April 2024, using  Spectral Imaging of the Coronal Environment (SPICE) to derive the sulfur-to-nitrogen ratio. EUV imaging and magnetograms provided additional context. Data were processed with the open-source Python tool Spectral Analysis Fitting Framework and Reduction of Noise (SAFFRON).
Both plumes showed sulfur fractionation that remained constant within uncertainties. The fractionated plasma was co-located with strong magnetic footpoints, in contrast with the surrounding interplume plasma. These results provide the evidence for sulfur fractionation in plumes and suggest, consistent with the ponderomotive force model, wave dynamics in the chromosphere as a driver.

\end{abstract}

\section{Introduction} \label{sec:Intro}

Coronal plumes are elongated, bright, and relatively narrow structures that extend from the solar surface outward through \myac{CHs}, either polar or equatorial. They are observed in visible, \myac{EUV}, and soft X-ray wavelengths \citep{plume_discription_1,plume_discription_2}, and their structural coherence can persist up to 30\,$\mathrm{R_\odot}$ \citep{plume_extension}. These features are embedded in regions of open magnetic field and are hypothesized to play a role in shaping the solar corona and contributing to the solar wind, especially its fast component. Understanding their thermodynamics, dynamics, and formation mechanisms is crucial for modeling the mass and energy transport in the outer solar atmosphere.

Coronal plumes are known to be denser and cooler than the interplume region in the surrounding coronal holes \citep{plume_dentemp_1,plume_dentemp_2}. They also have a relatively narrow \myac{DEM} profile in the corona, with emission predominantly in the range of 0.6--1.5\,MK \citep{Plume_FIP_3}, typically peaking near 0.8\,MK, slightly cooler than the ambient coronal hole plasma at the same height \citep{plume_living_rev}.

In addition to their thermal properties, coronal plumes have been the subject of interest in composition diagnostics. Comparing their elemental composition to in-situ solar wind measurements allows us to investigate the connectivity and mass origin of the heliospheric plasma. \myac{EUV} spectroscopy enables abundance diagnostics, which in many coronal structures reveal a so-called \myac{FIP} effect: an enhancement (or fractionation) of low-FIP elements ($\mathrm{FIP < 10\,eV}$; e.g., K, Ca, Fe, Si) in the corona relative to the photosphere \citep{1963ApJ...137..945P, Mayer_Atmosphere_FIP}. In contrast, high-FIP elements (e.g., O, Ne, Ar) remain largely unfractionated.

The FIP bias is quantified as:
\begin{equation}
\FIPrepr_\textrm{X} = \frac{Ab^C_\textrm{X}}{Ab^{Ph}_\textrm{X}},
\end{equation}
where $Ab^j_\textrm{X}$ is the abundance of element X in the corona (C) or photosphere (Ph). With \myac{EUV} observations, only relative abundance ratios between different elements X and Y are measurable:
\begin{equation}
\FIPrepr_{X/Y} = \frac{(Ab^C_X / Ab^C_Y)}{(Ab^{Ph}_X / Ab^{Ph}_Y)}.
\end{equation}
Assuming Y is a high-FIP element which has photospheric abundance ($\FIPrepr_Y = 1$), the fractionation ratio between X and Y is directly the FIP bias of X ($\FIPrepr_X = \FIPrepr_{X/Y}$). 

Within \myac{CHs}, the plasma composition is generally photospheric \citep{first_Si2S_EIS,CH_composition}, consistent with measurements in the fast solar wind \citep{SW_composition,SW_composition_1,SW_composition_2}, which originates from these regions. In contrast, enhanced FIP bias has been measured at \myac{CH} boundaries \citep{CHB_composition}. These locations are believed to contribute to the slow solar wind through interchange reconnection, where closed loops outside the \myac{CHs} containing fractionated plasma reconnect with open field lines, releasing the plasma into the heliosphere. While the slow solar wind is also fractionated, its composition is more variable, likely reflecting contributions from additional sources such as \myac{QS} regions \citep{QS_composition}, streamer belts \citep{Streamer_belt}, or pseudostreamers \citep{Pseudostreamers}.

Although the overall composition within \myac{CHs} is photospheric, small-scale structures embedded within them can display variable FIP bias, and plumes are a prime example. The literature on plume composition remains scarce, and the few existing studies report some contrasting results. \cite{Plume_FIP_1} used spectral data from the \myac{SUMER} spectrometer onboard the \myac{SOHO} and reported an enhanced FIP bias in plumes based on Mg/Ne diagnostics. Using \myac{CDS} on \myac{SOHO}, \citet{Plume_FIP_2} measured the Mg/Ne ratio and reported an overabundance of about 1.5. However, a later study that incorporated a more accurate \myac{DEM} profile for plumes questioned this interpretation \citep{Plume_FIP_3}, since the characteristic narrow \myac{DEM} profile of plumes can render Mg/Ne FIP bias diagnostics temperature sensitive due to the limitations of the 2-line ratio method for FIP diagnostics. More recently, \cite{Plume_FIP_S} used \myac{the EIS} onboard \myac{Hinode} to analyze the Si/S abundance ratio using the \myac{DEM} inversion method and found evidence that silicon is fractionated in plumes. They proposed that the FIP bias might evolve over a plume’s lifetime, potentially explaining the differences between their results and those of \cite{Plume_FIP_3}.

The composition variability across the solar surface is a potential valuable diagnostics for assessing solar wind connectivity. By comparing in-situ composition measurements with those obtained in the source regions, it is possible to constrain the origin of solar wind streams more precisely \citep{CON_EIS_3,EIS_ACE_Con,Yardley2023,Yardley2024}. However, the effectiveness of composition diagnostics for connectivity depends on three factors: a solid understanding of the underlying fractionation physics, the quality of both in-situ and remote-sensing data, and improved solar wind modeling. Progress in all of these areas continues to benefit from ongoing advancements.

On the theoretical side, the ponderomotive force model \citep{2009ApJ...695..954L, model_PMV_2015, Laming_S_Diag} is among the most successful frameworks for explaining FIP bias. In this model, Alfv\'en waves refract in the chromospheric density gradient, exerting a force preferentially on ions (mostly low-FIP elements), pushing them upward into the corona. The intermediate-FIP (mid-FIP) elements, such as sulfur, display more complex behavior, behaving sometimes as high-FIP or low-FIP elements depending on where the fractionation occurs.

\cite{Laming_S_Diag} showed theoretically that sulfur is more strongly fractionated when the ponderomotive force acts in the mid-chromosphere, a condition favored by leaking Alfv\'en waves from the corona, chromospheric reconnection, or wave conversion from the photosphere. 
Observationally, \citet{Sulfur_SEP} reported that sulfur, along with other mid-FIP elements such as phosphorus and carbon, exhibits variable fractionation. 
In the fast solar wind, sulfur is generally close to photospheric values, with possible mild enhancements of order $1.1$--$1.3$ \citep{Laming_S_Diag}. 
Larger values, up to $\sim 2$, occur mainly in the slow solar wind and in energetic particle populations, but even in those cases sulfur remains less enriched than low-FIP species such as Fe or Si. 
The long-term analysis of \citet{Alterman2025} further shows that sulfur in the slow solar wind varies on solar-cycle timescales, reaching higher values during periods of increased activity.
In particular, sulfur is more strongly enhanced in energetic particles from co-rotating interaction regions, whose sources are associated with slow solar wind streams, compared to gradual \myac{SEP} events thought to originate in closed coronal loops. 
This behaviour indicates that sulfur fractionation is sensitive to both the magnetic configuration and the wave environment of its source regions, making plume footpoints---where mixed-polarity fields and frequent reconnection events are observed \citep{plume_energy_release, high_res_plume}---likely sites for sulfur enhancement.

In this study, we are interested in the behavior of sulfur (mid-FIP element, 10.36\,eV) inside coronal plumes, with a particular focus on the time evolution of the FIP bias during plume development and decay. To that end, we analyze the \myac{SPICE} data on board \myac{SOLO}, complemented by the \solo/\myac{EUI}/\myac{FSI} imaging and the \solo/\myac{PHI}/\myac{HRT} magnetograms. In Sec.~\ref{sec:context}, we present an overview of the observed region and introduce the instruments used in this study. We then outline the methodology used to process the spectral imaging data, with a focus on the techniques applied to the \myac{SPICE} instrument (Sec.~\ref{sec:methods}). Here, we also introduce our open-source data processing tool (\texttt{SAFFRON}\footnote{Spectral Analysis, Fitting Framework, Reduction Of Noise Python module, \url{https://github.com/slimguat/saffron-spice}}), developed to clean and fit SPICE data for radiance and abundance map generation. The results, including the temporal and spatial evolution of the plasma composition and related observables, are presented in Sec.~\ref{sec:results}. A detailed discussion of the implications and interpretations as well as conclusions of this study are provided in Sec.~\ref{sec:discussions}.

\begin{figure}[H]
    \centering

    \begin{subfigure}[h]{0.48\textwidth}
        \centering
        \includegraphics[width= 2.5in,height=5.5in]{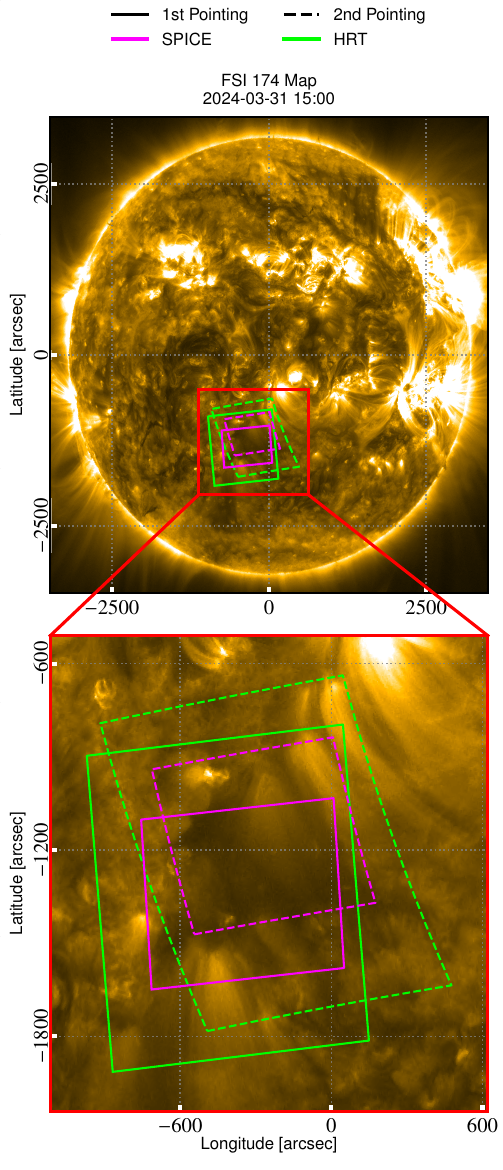}
        \caption{
            Context of the observations used in this study, showing the pointing coverage of \myac{SPICE} (magenta) and \myac{PHI}/\myac{HRT} (lime) during the two observation intervals: \mydateapj[]{2024,3,31} to \mydateapj[\noyear]{2024,4,01} for the first pointing (solid), and \mydateapj[]{2024,4,2--3} for the second pointing (dashed), overlaid on the \FSIFe map from \mydateapj[]{2024,3,31,15,00\,UT}.
            \textbf{Top:} Full-disk \FSIFe image with overlaid pointing contours. The red box indicates the zoomed-in region shown below.
            \textbf{Bottom:} Zoomed view of the target \myac{CH}.
        }
        \label{fig:context}
    \end{subfigure}
    \hfill
    \begin{subfigure}[h]{0.48\textwidth}
        \centering
        \includegraphics[scale=0.7]{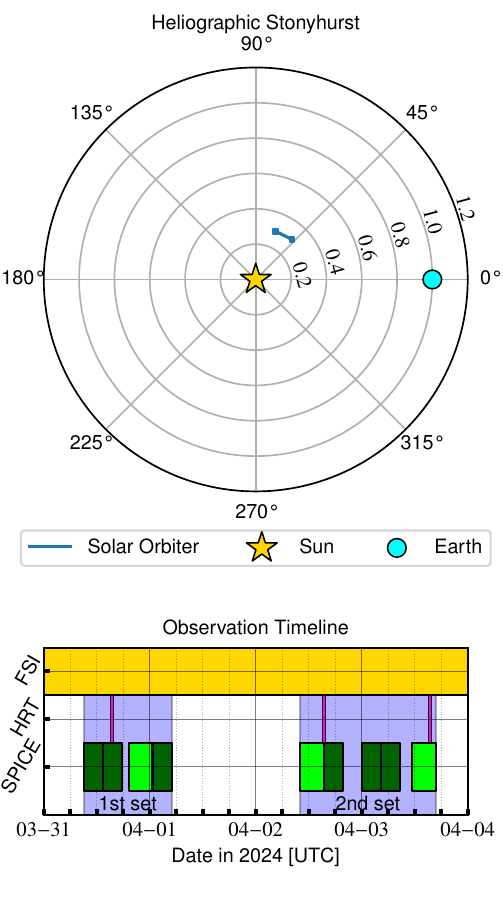}
        \caption{
            \solo position (top) and observation timeline (bottom) for the three \solo instruments used in this study between \mydateapj[]{2024,3,31} and \mydateapj[]{2024,4,4}. In the timeline panel, horizontal bands mark data acquisition intervals for \myac{FSI} (yellow), \myac{HRT} (magenta), and each \myac{SPICE} raster scan (dark green: 6\arcsec\ slit; lime: 4\arcsec\ slit). Shaded blue areas indicate the two main sets of \myac{CH} observations. In the top panel, the Sun is at the center (yellow star), the Earth’s position is at 1\,AU, 0$^\circ$ (light blue circle), and the \solo trajectory is shown in blue, with square markers denoting the first and last rasters (from 0.30\,AU to 0.29\,AU).
        }
        \label{fig:timeline}
    \end{subfigure}

    \caption{Observation context and timeline. (\subref{fig:context}) Context images and pointing. (\subref{fig:timeline}) \solo's relative locations and instrument timeline.}
    \label{fig:context_timeline}
\end{figure}
\newpage
\section{Observational Context} \label{sec:context}

\subsection{Target Description}
\label{subsec:CH_description}

The observations analyzed in this study were conducted during the \solo \myac{RSW}\,14 as part of the Fast Wind \myac{SOOP}\footnote{\url{https://www.cosmos.esa.int/web/solar-orbiter/soops-summary}}, which ran from \mydateapj[]{2024,3,31,09,00\,UT}
to \mydateapj[]{2024,4,3,21,00\,UT}. This \myac{SOOP} also included a second observing window during \myac{RSW}\,15, from \mydateapj[]{2024,4,12,01,00} to \mydateapj[\noyear]{2024,4,16,06,23\,UT}. 

The main objective of the Fast Wind \myac{SOOP} is to investigate the origins of the fast solar wind, primarily associated with open-field regions such as \myac{CHs}, and to connect remote-sensing observations of these sources to in-situ plasma and magnetic field measurements from \solo or other spacecraft.

Although four distinct targets were observed during the two \myac{RSWs} of the 2024 March/April Fast Wind \myac{SOOP} campaign, this study focuses exclusively on the first target: a well-defined equatorial \myac{CH}, observed in two intervals—from \mydateapj[]{2024,3,31,09,05,32} to \mydateapj[]{2024,4,1,04,57,10}, and from \mydateapj[]{2024,4,2,10,00,32} to \mydateapj[]{2024,4,3,16,48,19}. We refer to these periods as the ``first'' and ``second'' sets of observations, respectively. Figure~\ref{fig:context} displays an \FSIFe\ context image from \mydateapj[\noyear]{2024,3,31,15,00\,UT}, with a zoomed view of the targeted \myac{CH} shown in the bottom panel. During the observations of the target \myac{CH}, \solo was located at a distance of $\sim$0.30\,AU\ from the Sun, with a heliographic longitude between $47^{\circ}$ and $67^{\circ}$ relative to Earth (see Fig.\,~\ref{fig:timeline}-top). Coordinated observations were conducted with several instruments onboard \solo, all targeting the same \myac{CH} during overlapping intervals.

\subsection{Instrumentation and Data Products}
\label{subsec:inst_data}

This study uses the coordinated \solo observations combining EUV imaging, spectrophotometry, and EUV \added{spectroscopy}. The \myac{EUI}/\myac{FSI} 174\,\AA\ channel provides full-disk images dominated by \ion{Fe}{9}--\ion{Fe}{10} emission near 1\,MK with a 10\,min cadence, used to follow the evolution of the targeted regions. Daily \myac{PHI}/\myac{HRT} \myac{LOS} magnetograms (15:00\,UT; see bottom panel of Fig.\,\ref{fig:timeline}) supply the photospheric flux information and are co-aligned with \FSIFe\ for contextual analysis. Composition rasters from \myac{SPICE} were acquired with the 4\arcsec\ and 6\arcsec\ slits (120\,s exposures, dense rasters), producing 4--5\,h scans. The two \solo pointings were slightly offset, producing partial overlap in the narrow-\myac{FOV} instruments (Fig.\,\ref{fig:context}); \myac{HRT} covered the full target \myac{CH} in both cases, while \myac{SPICE} sampled only part of the region, with an overlap of 56--59\% between the two rasters.

The bottom panel of Fig.\,\ref{fig:timeline} shows the \myac{SPICE} observation timeline. Dark green bars indicate raster scans with the 6\arcsec\ slit and lime bars indicate scans with the 4\arcsec\ slit. The number of scan steps was adjusted to keep approximately the same FOV in both cases. Four rasters were obtained during the first observation set and five during the second. The SPICE ``composition'' rasters include key emission lines from the transition region and low corona, used to infer elemental abundances. Details of the line selection and analysis are given in Sec.~\ref{sec:methods}. In this study we exploit the L2 data from the official SPICE data \release{\footnote{\url{https://spice.osups.universite-paris-saclay.fr/spice-data/release-5.0/release-notes.html}}}\ \citep{spice_data_release_5}.

\subsection{Plume Identification and Evolution}
\label{subsec:plume_evolution}

\begin{figure}[t]
    \centering

    \begin{subfigure}[h]{0.48\textwidth}
        \centering
        \includegraphics[scale=0.7]{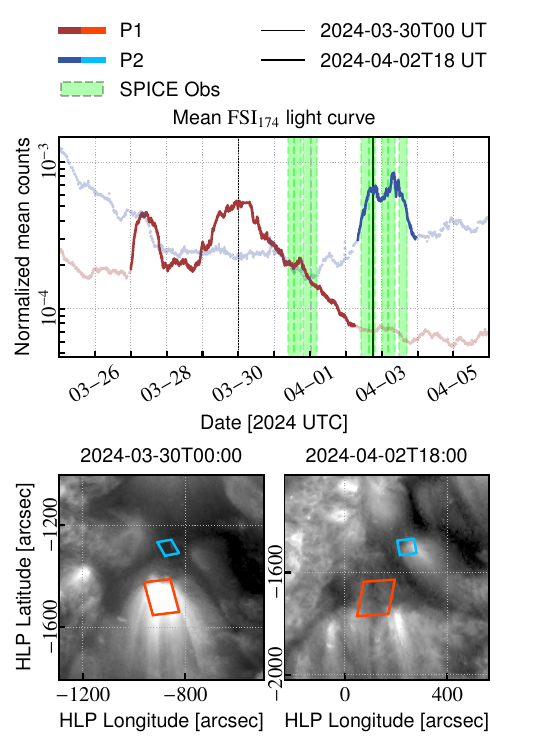}
        \caption{
            The top panel shows the normalized mean intensity evolution from \mydateapj{2024,3,25} to \mydateapj{2024,4,6}, measured in counts per second by \FSIFe, for two selected footpoint regions ``P1'' (red) and ``P2'' (blue). These regions correspond to the bases of Plume~1 and Plume~2, respectively, and are used consistently throughout this study. The red and blue lines are stronger over the durations in which each plume was present. The lime-colored vertical stripes indicate the times and durations of the nine SPICE raster scans targeting the same region. The bottom panels show \FSIFe context maps, in helioprojective coordinates, during the lifetimes of each plume: the bottom-left panel corresponds to Plume~1 (\mydateapj[\noyear]{2024,3,30,00,00\,UT}), and the bottom-right to Plume~2 (\mydateapj[\noyear]{2024,4,2,18,00\,UT}), with the selected regions outlined in orange and cyan.
        }
        \label{fig:plume_light_curve}
    \end{subfigure}
    \hfill
    \begin{subfigure}[h]{0.48\textwidth}
        \centering
        \includegraphics[scale=0.7]{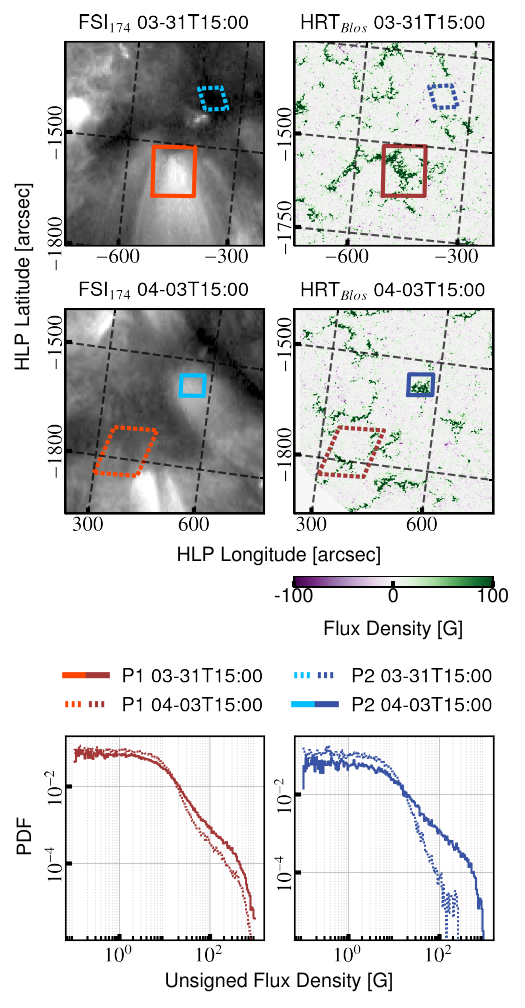}
        \caption{
            Magnetic and intensity context for the ``P1'' and ``P2'' region selections introduced in Fig.\,\ref{fig:plume_light_curve}. The selections are shown on the \FSIFe intensity maps (left panels) and the \PHIBLOS magnetograms (right panels), for both SPICE observation sets (top and bottom rows). The histogram panel shows the probability distribution function (PDF) of the unsigned magnetic flux density within the P1 (red/orange) and P2 (blue/cyan) selections, based on \PHIBLOS data. Solid lines indicate the time when the corresponding plume is present; dashed lines correspond to times when the plume is not present.
          }
        \label{fig:phi_maps}
    \end{subfigure}

    \caption{(\subref{fig:plume_light_curve}) Intensity evolution and plume context for Plume~1 and Plume~2. (\subref{fig:phi_maps}) Magnetic and intensity context of plume footpoints with flux density distributions.}
    \label{fig:plume_combined}
\end{figure}

During the observations, we identified two plumes within the target \myac{CH} (see Fig.~\ref{fig:plume_light_curve}). These structures were tracked using normalized \FSIFe light curves, obtained following the processing steps described in App.~\ref{sec:app___light_curves_calculations}. Two fixed regions were selected in the Carrington frame and are used consistently throughout this study; we refer to them as ``P1'' and ``P2'', the locations in which Plume~1 and Plume~2 formed. Plume~1 was present from \mydateapj[\noyear]{2024,3,27,00,00\,UT} to \mydateapj[\noyear]{2024,4,2,06,00\,UT}, lasting about 6 days and 6 hours. Plume~2 developed later, from \mydateapj[\noyear]{2024,4,2,08,00\,UT} to \mydateapj[\noyear]{2024,4,3,23,00\,UT}, with a duration of about 1 day and 15 hours. These intervals were estimated from visual inspection of the \FSIFe images and time series, with an uncertainty of roughly 2\,hours. Throughout the observing period, Plume~1 remained brighter and more extended than Plume~2. The first set of \myac{SPICE} rasters, obtained on \mydateapj[\noyear]{2024,3,31} and early \mydateapj[\noyear]{2024,4,1}, sampled Plume~1 during its decay. The second set, taken about one day later, covered Plume~2 shortly after its formation and throughout most of its evolution. The timing of all nine rasters is shown in Fig.~\ref{fig:plume_light_curve}.

The magnetic footpoints of both regions are shown in Fig.~\ref{fig:phi_maps}, together with the corresponding distributions of the unsigned magnetic flux density. Values below 10\,G are dominated by noise (\citealt{HRT_noise} estimate a noise level of about 8.3\,G under \myac{QS} conditions), so only the behaviour above this threshold is meaningful. In both regions, the flux declines beyond 10\,G, and during plume activity the distributions extend further toward higher field strengths, with contributions above 50\,G marking the concentrated magnetic flux at the footpoints. In ``P1'', this high-field component remains visible after the decay of Plume~1, although with reduced amplitude, and its magnetic footprint is broader than that of Plume~2. When restricting the analysis to pixels above 50\,G, the average unsigned magnetic flux density during plume activity is around 220\,G in both regions.

\section{Methods} \label{sec:methods}

\begin{table*}[t]
    \caption{\added{Spectral lines in the SPICE composition rasters (wavelength in \AA) with their transitions and peak formation temperatures ($\log T$).} Blended lines are marked with *.}
    \centering
    \label{tab:spectral_lines}
    \begin{tabular}{p{0.2\textwidth} p{0.3\textwidth} p{0.07\textwidth}}
    \toprule
        Line & Transition & \added{$\log T$} \\
    \midrule
        \ion{Ar}{8} 700.240* & \added{3s $^2$S$_{1/2}$\,--\,3p $^2$P$_{3/2}$} & \added{5.65} \\

        \ion{O}{3} 702.337* & \added{2s$^2$2p$^2$ $^3$P$_0$\,--\,2s2p$^3$ $^3$P$_1$} & \added{4.91} \\
        \ion{O}{3} 702.838* & \added{2s$^2$2p$^2$ $^3$P$_1$\,--\,2s2p$^3$ $^3$P$_0$} & \added{4.91} \\
        \ion{O}{3} 702.896* & \added{2s$^2$2p$^2$ $^3$P$_1$\,--\,2s2p$^3$ $^3$P$_1$} & \added{4.91} \\
        \ion{O}{3} 703.855* & \added{2s$^2$2p$^2$ $^3$P$_2$\,--\,2s2p$^3$ $^3$P$_2$} & \added{4.91} \\

        \ion{Mg}{9} 706.060 & \added{2s$^2$ $^1$S$_0$\,--\,2s2p $^3$P$_1$} & \added{6.00} \\

        \ion{S}{4} 748.393 & \added{3s$^2$3p $^2$P$_{1/2}$\,--\,3s3p$^2$ $^2$P$_{1/2}$} & \added{4.95} \\
        \ion{Mg}{9} 749.552* & \added{2s2p $^1$P$_1$\,--\,2p$^2$ $^1$D$_2$} & \added{6.00} \\
        \ion{S}{4} 750.221* & \added{3s$^2$3p $^2$P$_{3/2}$\,--\,3s3p$^2$ $^2$P$_{3/2}$} & \added{4.95} \\

        \ion{O}{5} 760.446* & \added{2s2p $^3$P$_2$\,--\,2p$^2$ $^3$P$_2$} & \added{5.32} \\
        \ion{O}{5} 762.004* & \added{2s2p $^3$P$_2$\,--\,2p$^2$ $^3$P$_1$} & \added{5.32} \\

        \ion{N}{4} 765.152 & \added{2s$^2$ $^1$S$_0$\,--\,2s2p $^1$P$_1$} & \added{5.10} \\
        \ion{Ne}{8} 770.428 & \added{1s$^2$2s $^2$S$_{1/2}$\,--\,1s$^2$2p $^2$P$_{3/2}$} & \added{5.80} \\
        \ion{Mg}{8} 772.260 & \added{2s$^2$2p $^2$P$_{3/2}$\,--\,2s2p$^2$ $^4$P$_{5/2}$} & \added{5.90} \\

        \ion{S}{5} 786.468* & \added{3s$^2$ $^1$S$_0$\,--\,3s3p $^1$P$_1$} & \added{5.15} \\
        \ion{O}{4} 787.710* & \added{2s$^2$2p $^2$P$_{1/2}$\,--\,2s2p$^2$ $^2$D$_{3/2}$} & \added{5.12} \\

        \ion{H}{1} 972.537 & \added{1s $^2$S$_{1/2}$\,--\,4p $^2$P$_{3/2}$} & \added{4.25} \\

        \ion{Na}{6} 988.709* & \added{2s$^2$2p$^2$ $^3$P$_2$\,--\,2s2p$^3$ $^5$S$_2$} & \added{5.60} \\
        \ion{N}{3} 989.799* & \added{2s$^2$2p $^2$P$_{1/2}$\,--\,2s2p$^2$ $^2$D$_{3/2}$} & \added{4.85} \\
        \ion{N}{3} 991.577* & \added{2s$^2$2p $^2$P$_{3/2}$\,--\,2s2p$^2$ $^2$D$_{5/2}$} & \added{4.85} \\
        \ion{Ne}{6} 992.683* & \added{2s$^2$2p $^2$P$_{1/2}$\,--\,2s2p$^2$ $^4$P$_{3/2}$} & \added{5.53} \\
    \bottomrule
    \end{tabular}
\end{table*}

%
%
%

To derive plasma parameters, we introduce the Python module \myac[\acLong]{SAFFRON} developed for \myac{SPICE} data analysis. This open-source tool encompasses multiple stages of processing, including data cleaning (e.g., despiking and cosmic ray removal, data binning), uncertainty estimation, spectral fitting, and post-processing tasks such as FIP bias diagnostics. It is designed to support both rapid inspection and in-depth analysis through in-depth fine tuning.

\myac{SPICE} data preparation included cosmic ray removal using a sigma-clipping method and binning to improve the \myac{SNR} (which results in $\sim$12\arcsec resolution in both directions). Spectral fitting was performed on both the original and binned datasets. The unbinned data retained the full spatial resolution and were used for identifying morphological structures, while the binned data provided more reliable quantitative results for composition diagnostics such as FIP bias. 
Example radiance maps illustrating this difference between full-resolution and binned data are shown in Fig.~\ref{fig:spice_rad_sample}.
The full description of the binning strategy, the fitting model, and the associated implementation is presented in App.~\ref{sec:app__spice_processing}, together with the centroid constraints applied to weak or blended lines to ensure reliable fitting of the key diagnostic lines.
 
\begin{figure*}
    \centering
    \includegraphics[width=1\linewidth]{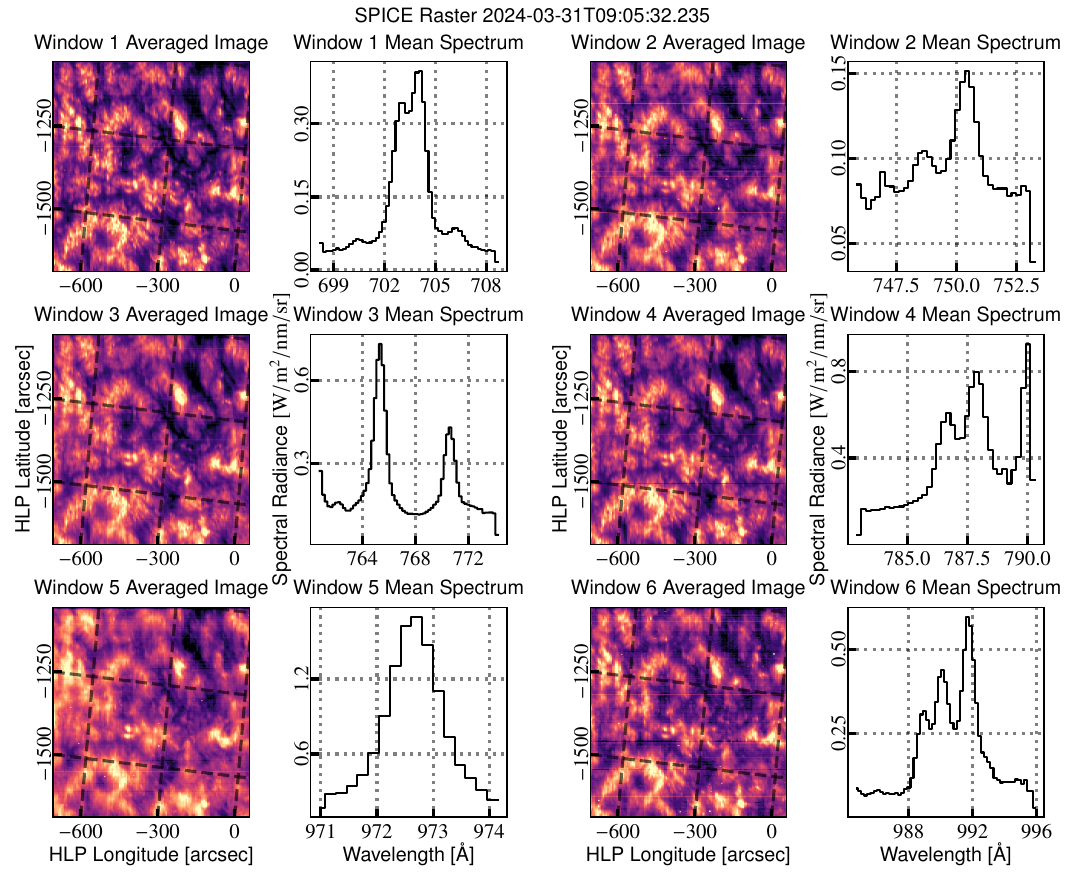}
    \caption{
        Sample SPICE composition raster from \mydateapj[]{2024,3,31,09,05,32\,UT}. First and third columns 
        show mean images for each of the six spectral windows (window~1 to window~6), averaged along wavelength dimension. Right to each image panel we
        show the corresponding average spectra per window across the spatial axis.
        }
    \label{fig:spice_sample}
\end{figure*}

The full list of fitted spectral lines is given in Table~\ref{tab:spectral_lines}. Also a sample of the raster windows is given in Fig.\,\ref{fig:spice_sample} where we show the wavelength-averaged image and the corresponding average spectra. 

\begin{figure}[t]
    \centering
    \includegraphics[width=0.8\linewidth]{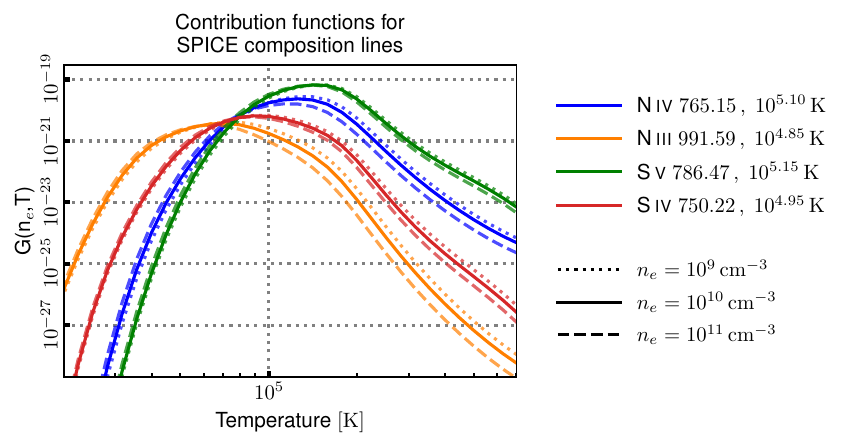}
    \caption{
    Contribution functions (in $\mathrm{erg\,cm^{3}\,s^{-1}\,sr^{-1}}$) for the \myac{SPICE} composition lines used in this study, computed with CHIANTI~v.11 (using the IDL package of CHIANTI)
    \added{for three constant electron densities of $10^{9}$, $10^{10}$, and $10^{11}\,\mathrm{cm^{-3}}$}.
    Each curve is labeled with the ion, wavelength \added{(in~\AA)}, and $T_{\mathrm{max}}$, the peak formation temperature, in~K.
    }

    \label{fig:gofnt}
\end{figure}

FIP bias maps were constructed using the \myac[\acLong]{LCR}, called from \myac{SAFFRON} via the \texttt{fiplcr} module:
\begin{align}
    \FIPrepr = \frac{
    \sum\limits_{i \in (\mathrm{LF})} {\alpha_i} \cdot I_i /{A_i^P}
    }{
    \sum\limits_{j \in (\mathrm{HF})} {\beta_j}  \cdot I_j/{A_j^P}
    }.
    \label{FIP_equation}
\end{align}
Here, $\alpha_i$ and $\beta_j$ are coefficients for low- and high-FIP lines, and $A^P$ denotes photospheric abundances obtained by \cite{photospheric_composition}. The atomic physics data are obtained using Chianti atomic database \citep[version 11.0.2,][]{CHIANTI_paper,chianti11}.
HF denotes the lines of the elements considered as high-FIP, while LF is for the lines of elements considered as low-FIP.

These coefficients were optimized using the CHIANTI-provided typical DEMs for active regions, coronal holes, and quiet Sun. While \cite{LCR} found that the \myac{LCR} method is relatively insensitive to the DEM choice, it requires that the selected lines have sufficiently similar contribution functions to allow the cost function (Eq.\,21 in \cite{LCR}) to reach a well-defined minimum.

The selected lines from a high-FIP element are \ion{N}{4}~765.152\,\AA\ and \ion{N}{3}~991.577\,\AA, and the lines from a lower-FIP element are \ion{S}{4}~750.221\,\AA\ and \ion{S}{5}~786.468\,\AA. Their contribution functions peak around 0.08--0.15\,MK, as shown in Fig.\,\ref{fig:gofnt}, corresponding to transition region temperatures. 
All of the selected spectral lines are considered optically thin including \ion{N}{3}~991.577\,\AA. Previous studies have shown that this line with a low formation temperature remains optically thin in coronal holes including plume footpoints \citep{Jordan2001,SPICE_Mosaic}. 
\added{A dedicated analysis of the S/N composition diagnostics with \myac{SPICE}, also part of this special issue \citep{ZambranaPrado2025}, further assesses the performance of the LCR method for these line ratios.}

To apply the \myac{LCR} method to the \myac{SPICE} data, we adopted a fixed electron density of $10^{10}\,\mathrm{cm^{-3}}$. 
In principle, density diagnostics are available in the \myac{SPICE} range through the \ion{O}{5} 761.128\,\AA/760.446\,\AA\ ratio. 
However, this interval also contains three nearby \ion{O}{5} lines that form a blended group, and in our dataset this part of the spectrum falls at the edge of the spectral window (see Fig.\,\ref{fig:spice_sample}, Window~3 mean spectrum). 
The full set of lines is therefore not captured, making the \ion{O}{5} ratio unreliable for density estimation.

A value of $10^{10}\,\mathrm{cm^{-3}}$ is an upper bound for typical transition-region conditions in \myac{CHs} \citep{density_CH_TR}. 
Since higher assumed densities reduce the inferred FIP bias, this choice results in a conservative estimate; any measured enhancement above the uncertainties can be regarded as real and possibly stronger than reported. 
Density-related considerations are discussed further in App.~\ref{sec:app___density_effects}.

To improve the \myac{SNR}, composition maps were generated from binned data, and pixels with relative uncertainties exceeding 50\% were excluded from the FIP bias analysis. 
Here, the relative uncertainty refers to the ratio of the propagated intensity error to the fitted line intensity, where the intensity errors are obtained by propagating uncertainties arising from photon statistics and from the instrument’s broad \myac{PSF}, which also deviates slightly from a Gaussian profile. 
These effects dominate over other sources of error, such as uncertainties in atomic data \citep[from CHIANTI v.11,][]{chianti11} or radiometric calibration \citep{spice_data_release_5}. 
The resulting relative uncertainties on the fitted line intensities directly provide the error estimates on the derived FIP bias values. 
The full uncertainty calculation and propagation procedure is described in App.~\ref{sec:app__errors}.


\section{Results} \label{sec:results}

\begin{figure*}
    \centering
    \includegraphics[width=\linewidth]{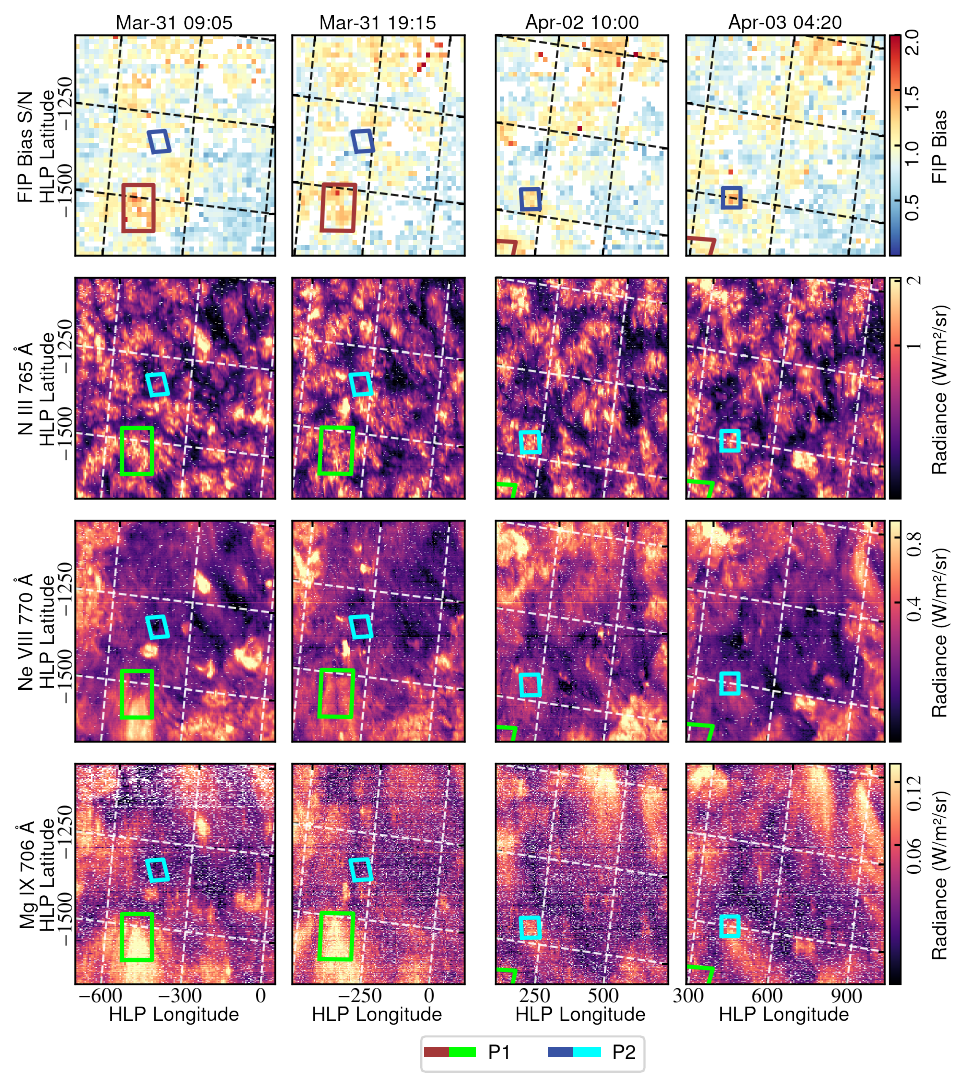}
    \caption{Sample of SPICE observations from each observation set: the first set is shown in the first two columns, and the second set in the last two columns. Each column displays, from top to bottom, the S/N FIP bias maps and radiance maps for SPICE~\ion{N}{3} (used in the FIP bias calculation; low-\myac{TR}), \ion{Ne}{8} (top-\myac{TR}), and \ion{Mg}{9} (corona). The regions labeled ``P1'' (red in the FIP bias maps and green in the radiance maps) and ``P2'' (blue in the FIP bias maps and light blue in the radiance maps) indicate the locations of Plume~1 and Plume~2, respectively. These regions were selected based on visual identification of the plume footpoints in one map and propagated to the others by fixing the Carrington coordinates of the region vertices.
}
    \label{fig:P12_maps}
\end{figure*}

\subsection{FIP bias within plumes}
Figure~\ref{fig:P12_maps} shows the resulting FIP bias maps (first row) and selected SPICE radiance maps (subsequent rows) for four selected raster scans, one per column. The first two scans were acquired during the \mydateapj[\noyear]{2024,3,31} pointing, and the last two during the \mydateapj[\noyear]{2024,4,2--3} pointing. The radiance maps are ordered by peak emission temperature: \ion{N}{3} 765.152\,\AA\ (0.15\,MK), \ion{Ne}{8} 770.428\,\AA\ (0.63\,MK), and \ion{Mg}{9} 706\,\AA\ (1.00\,MK). The FIP bias maps trace the plasma composition between 0.1\,MK and 0.15\,MK.

The FIP bias across the full \myac{SPICE} \myac{FOV} ranges between 0.8 and 1.0, indicating that sulfur remains close to its photospheric abundance. However, the ``P1'' region in Fig.\,\ref{fig:P12_maps}, corresponding to the location of Plume~1 footpoint, shows consistent fractionation during the first observation set (first and second columns), with FIP bias values between 1.1 and 1.5. During these observations, Plume~1 is clearly visible in the \ion{Mg}{9} maps. In the second observation set (third and fourth columns), ``P1'' lies mostly outside the SPICE \myac{FOV}, and is therefore excluded from further analysis.

The ``P2'' region shows fractionation only during the second observation set (starting from \mydateapj[\noyear]{2024,04,2}), corresponding to the presence of Plume~2. The plume is then visible in \ion{Mg}{9} maps. In contrast, during the first set—prior to the appearance of Plume~2—the FIP bias in this region is approximately 1.0, consistent with photospheric sulfur abundance. In both plumes, the fractionated areas align with enhanced \ion{N}{3} emission in the plume footpoints.

Although Plume~1 was observed (during the first set of observations) in its decay phase, its structure remains clearly detectable in \ion{Mg}{9}, as shown in the fourth row of Fig.\,\ref{fig:P12_maps}. The \ion{Ne}{8} maps, however, reveal temporal variations in the emission from Plume~1 (first and second column of Fig.~\ref{fig:P12_maps}), with the weakest signal recorded on \mydateapj[\noyear]{2024,3,31,19:15}. This suggests thermal evolution of the plume structure at transition region temperatures. Plume~2 was observed throughout its lifetime but appears smaller and systematically fainter in \ion{Ne}{8}. A lower emission measure at this temperature may arise from a cooler thermal distribution and/or a reduced plasma density; both interpretations are consistent with Plume~2 being intrinsically weaker than Plume~1.


\subsection{In-depth Plume FIP bias analysis and error analysis}

The FIP bias enhancements observed in the ``P1'' and ``P2'' regions originate from the footpoints of Plume~1 and Plume~2. However, we focus solely on assessing the presence of sulfur fractionation rather than deriving precise FIP bias values, due to the lack of precise electron density diagnostics as detailed in Sec.~\ref{sec:methods}. 
Because of the assumed density, the reported FIP bias values should be regarded as conservative lower limits. Thus, whenever the measured FIP bias exceeds unity by more than its associated uncertainty, this indicates genuine sulfur enrichment.
Consequently, values near or below unity are interpreted as photospheric. In particular, we emphasize that the values below one are not indicative of an \myac{IFIP} effect, as they could be the outcome of the density assumption.




Figure~\ref{fig:P12_hist} illustrates the FIP bias measurements and associated uncertainties across the selected areas ``P1'' and ``P2'' during two sets of \myac{SPICE} observations. These distributions allow us to assess the spread in composition values and the reliability of the inferred FIP bias in each plume over time. To distinguish reliably fractionated pixels from those consistent with photospheric composition, we applied the criterion \(\FIPrepr - \Delta\FIPrepr/2 > 1\), where \(\FIPrepr\) is the measured FIP bias and \(\Delta\FIPrepr\) its uncertainty. Pixels satisfying this condition are represented in the green-shaded domain of the scatter plots and are considered confidently fractionated. The fraction of such pixels can be visually tracked across scans and directly compared between Plume~1 and Plume~2.


In the first observation set, the ``P1'' histograms show a clear population of fractionated pixels, with values typically around 1.2 and reaching up to 2.0. Conversely, the ``P2'' region displays a clear FIP bias enhancement during the second observation set when Plume~2 is present, with values reaching up to 1.5. No such signature is present in the first set, prior to the appearance of Plume~2.

These findings indicate that sulfur becomes fractionated in the presence of plumes, while it remains photospheric prior to the Plume~2 formation. 



\subsection{FIP bias evolution over time}


After confirming that the targeted plumes exhibit sulfur fractionation, we quantify the FIP bias evolution over time for both regions ``P1'' and ``P2''. To achieve this, we compute the weighted mean FIP bias within each region using the following expression:
\begin{equation}
\bar\FIPrepr = \frac{\sum_i^N \FIPrepr_i \cdot \frac{1}{\Delta\FIPrepr_i^2}}{\sum_i^N \frac{1}{\Delta\FIPrepr_i^2}},
\label{eq:mean_formula}
\end{equation}
where $\FIPrepr_i$ and $\Delta\FIPrepr_i$ denote the FIP bias and its estimated uncertainty for pixel $i$.

\begin{figure}[htbp]
    \centering

    \begin{subfigure}[t]{0.48\textwidth}
        \centering
        \includegraphics[width=\linewidth]{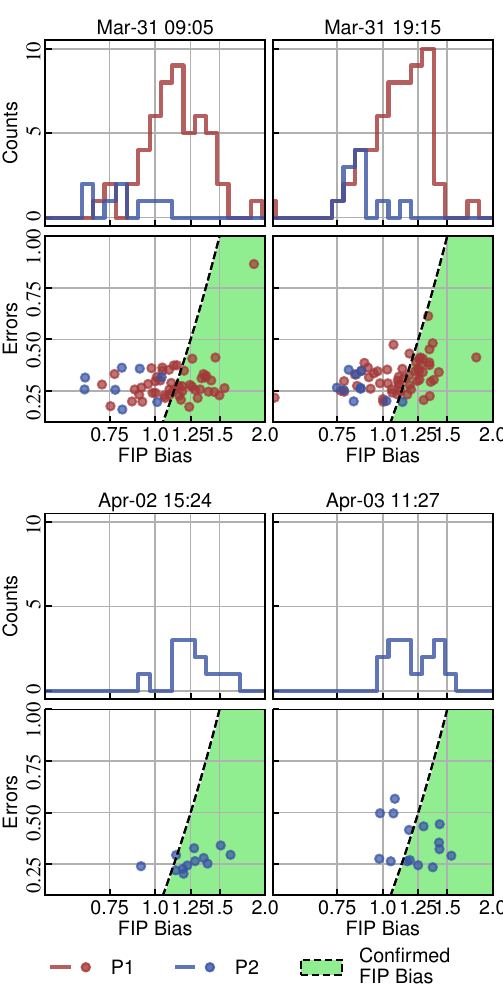}
        \caption{
            FIP bias distribution and uncertainty analysis for pixels in the plume footpoint regions ``P1'' (red) and ``P2'' (blue). 
            Each vertical pair of histogram and scatter plot corresponds to a raster scan selected as in Fig.\,\ref{fig:P12_maps}. 
            The top two pairs represent the first SPICE observation set (\mydateapj[\noyear]{2024,3,31}), and the bottom four pairs represent the second set (\mydateapj[\noyear]{2024,4,2--3}). 
            In each pair: \textbf{Top panel} shows the FIP bias histogram; \textbf{Bottom panel} shows the scatter plot of FIP bias versus uncertainty. 
            The dashed black line marks the threshold \( \FIPrepr - \Delta\FIPrepr/2 = 1 \); pixels in the green-shaded region are considered reliably fractionated.
        }
        \label{fig:P12_hist}
    \end{subfigure}
    \hfill
    \begin{subfigure}[t]{0.48\textwidth}
        \centering
        \includegraphics[width=\linewidth]{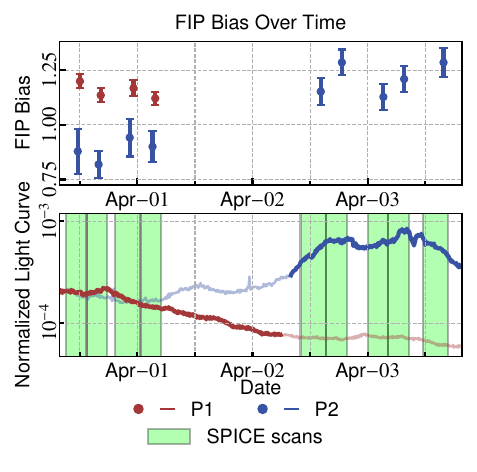}
        \caption{
            Temporal evolution of the FIP bias in regions ``P1'' (red) and ``P2'' (blue), corresponding to the footpoints of Plume~1 and Plume~2. 
            \textbf{Top panel:} Weighted mean FIP bias in each region, derived from the maps in Fig.\,\ref{fig:P12_maps} and the distributions in Fig.\,\ref{fig:P12_hist}, using Eq.\,\ref{eq:mean_formula}. 
            Each point is associated with the average scan time of the relevant sub-region. 
            \textbf{Bottom panel:} Normalized mean intensity evolution from \FSIFe during the SPICE observation period. Lime bars indicate the time windows of the SPICE rasters, as in Fig.\,\ref{fig:plume_light_curve}.
        }
        \label{fig:P1_FIPvsTime}
    \end{subfigure}

    \caption{
        (\subref{fig:P12_hist}) FIP bias distributions and uncertainty analysis for plume footpoint regions P1 and P2. 
        (\subref{fig:P1_FIPvsTime}) Temporal evolution of the weighted mean FIP bias and corresponding intensity variations in the same regions. 
    }
    \label{fig:P12_combined}
\end{figure}

Figure~\ref{fig:P1_FIPvsTime} presents the results. The top row shows the time evolution of the weighted mean FIP bias, while the bottom panels reproduce the normalized light curves from Fig.\,~\ref{fig:plume_light_curve}, now zoomed into the SPICE observation windows.

We find that the mean FIP bias during the lifetime of both plumes is around 1.2, occasionally reaching up to 1.3. Both regions show comparable fractionation behavior. Notably, the FIP bias in ``P2'' remains photospheric prior to the appearance of Plume~2 and six hours after its appearance, the average FIP bias indicates clear enrichment, and continues to increase further in the following observation, which occurs 10 hours after the plume’s appearance. The FIP bias then stabilizes until the final observation, which took place seven hours before the complete decay of the plume.


As previously mentioned, the assumed electron density of $10^{10}\,\mathrm{cm^{-3}}$ provides a conservative lower limit of the FIP bias. If we consider a broader range down to the lower \myac{CH} density of $10^9\,\mathrm{cm^{-3}}$, the actual FIP bias could be $\sim$20\% higher, bringing peak values close to 1.4 or more.

\section{Discussions and Conclusion} \label{sec:discussions}

Our results clearly indicate that sulfur undergoes fractionation inside the observed plumes. Both plumes show comparable overall FIP bias levels; however, the fractionated area is larger in Plume~1. This correlates with Plume~1 being more intense (Fig.\,\ref{fig:plume_light_curve}) and having a larger and stronger magnetic flux concentration (exceeding 100\,G; Fig.\,\ref{fig:phi_maps}).

Throughout the lifetime of each plume, no time-dependent variation of the FIP bias was detected --- at least not within the measurement uncertainties. However, a distinct increase in FIP bias is observed at the location of Plume~2 between its pre-formation phase and active phase. Between \mydateapj[\noyear]{2024,3,31}--\mydateapj[\noyear]{2024,4,1} and \mydateapj[\noyear]{2024,4,2--3}, the FIP bias increased from approximately 0.9 to 1.2, indicating that sulfur became fractionated within the six hours leading up to the first SPICE observations of the plume. Because we assume a relatively high electron density for a coronal hole environment in our analysis, the derived FIP bias values represent conservative lower limits. Accounting for this, the actual FIP bias in the plumes could be 10--20\,\% higher. A slight variation is observed in Plume~2 during its activity, but it remains almost within the uncertainties of the measurements.

This sulfur fractionation behavior can be interpreted with the ponderomotive force model \citep{2009ApJ...695..954L,model_PMV_2015}, in which the FIP bias is produced by the ponderomotive acceleration. This acceleration scales with the gradient of the transverse wave electric energy, $\partial \delta E_\perp^2 / \partial z$, and is directed upward, from the chromosphere towards the corona. Under such conditions, the ponderomotive force peaks near the top of the chromosphere. \citet{Laming_S_Diag} also demonstrated that mid-FIP fractionation, including sulfur, is possible if Alfv\'en waves penetrate into the mid-chromosphere. In their model, one source of these waves is coronal Alfv\'en waves that partially transmit through the density barrier of the \myac{TR}, reaching the upper chromosphere where the ponderomotive force is generated. Moreover, torsional Alfv\'enic waves can propagate deeper into the mid-chromosphere, thereby enabling mid-FIP fractionation. Alfv\'en waves may also be generated directly within the chromosphere, since reconnection can take place at multiple atmospheric heights \citep{Plume_reconnection}, or they may arise from upward-propagating waves originating in the photosphere through slow-to-Alfv\'en \citep{slowAlfvenConv} or fast-to-Alfv\'en \citep{fastAlfvenConv_1,fastAlfvenConv_2} mode conversion. 
Although Alfvén waves converted from p modes leaking into plume regions have been reported as quasi-periodic 3--5 minute activity in the corona \citep{plume_energy_release}, \citet{Chromo_corona_waves} showed that in quiescent conditions (non-flaring Sun), upward waves alone do not generate a significant ponderomotive force.

It is therefore reasonable, based on the predictions of the ponderomotive-force model, to expect plume plasma to undergo fractionation. 
Plumes are not single monolithic structures, but rather the collective outcome of numerous small-scale events, often referred to as plumelets \citep{plumelets,plume_energy_release}. Each plumelet forms through the reconnection of background open-field lines with small closed loops rooted in mixed-polarity magnetic patches, where the main dominant polarity interacts with parasitic inverse polarities, as found by \citet{high_res_plume} using high-resolution observations from BBSO/NIRIS. These reconnection events generate small-scale plasma jets, or jetlets \citep{jetlets,singular_jetlets}, which collectively sustain the large-scale plume structure \citep{Plume_reconnection}. The reconnection dynamics naturally excite Alfv\'enic perturbations. In this framework, the observed plume fractionation can be understood as a consequence of the ponderomotive force acting on plasma in the upper and mid-chromosphere. The same dynamics that sustain plume activity also supply the necessary Alfv\'enic amplitudes. These are absent in inter-plume regions, which explains why the \myac{CH} plasma generally lacks fractionation signatures \citep{CH_composition}. A fraction of these perturbations are torsional Alfv\'en waves, which can penetrate deeper into the chromosphere and provide a natural explanation for the sulfur enhancement that we observe.

In this study, we do not have direct access to wave diagnostics in the chromosphere, which prevents us from directly constraining the role of chromospheric wave activity in the observed fractionation. Specifically, we cannot quantify the wave energy, determine its latitudinal gradients, or assess its polarization. Nor can we confirm which processes act as the dominant sources of the waves. These limitations highlight the need for future observations with high spatial and temporal resolution, both in the chromosphere and the lower corona, to firmly establish the connection between chromospheric wave properties and fractionation signatures.

\citet{Plume_FIP_S} analyzed Si/S fractionation using Hinode/\myac{EIS} data and found silicon enhanced over sulfur by a factor of 1.1 to 1.3. However, because our results suggest sulfur itself is fractionated in plumes, their silicon enrichment might be underestimated. If sulfur is enhanced to the degree we observed, the true ratio of silicon to a typical low-FIP element (e.g., O, Ne, Ar) would be higher (up to 1.7). It is also possible that plume cases where the Si/S ratio is close to unity do not reflect an absence of silicon fractionation, but rather comparable fractionation of both silicon and sulfur.

That study also investigated temporal variability and found that the FIP bias remained constant over time for three out of four plumes, while one displayed a clear time-dependent evolution. Furthermore, plumes that decayed were found to revert to interplume (i.e., photospheric) composition --- a similar trend we observe in the FIP bias of region ``P2'' before (first set) and during (second set) Plume~2 activity. Although we detect no significant temporal evolution in the FIP bias of our studied plumes, such variability may depend on plume-specific properties that modulate the wave energy content and the generation of Alfvén waves, thereby influencing the fractionation rate. A larger statistical sample may be required to identify consistent trends. If FIP bias is indeed modulated by the occurrence and amplitude of Alfv\'en waves, then plumes with more dynamic evolution or changing magnetic configurations may exhibit stronger time variability. Notably, the most pronounced changes in fractionation must occur during the early phases of plume formation or during their decay (within a few hours at most).

Based on the sulfur fractionation observations in our results, together with the theoretical predictions of the ponderomotive-force model \citep{2009ApJ...695..954L, model_PMV_2015, Laming_S_Diag}, it is possible that the combination of mid-FIP and low-FIP abundance enhancements could serve as an indirect proxy for the presence and transport of Alfv\'en waves in the mid-chromosphere.. It might also enable direct linkage between in-plume composition and solar wind observations. However, in this study, we lack connectivity with in-situ instruments and therefore cannot establish such links \citep[][]{EIS_ACE_Con}.

To investigate the role of small-scale structures in FIP bias formation, coordinated multi-instrument observations with both high spatial and temporal resolution are essential. On one side, high-resolution (resolving down to plumelet scale, $\sim 100$\,km at large) \myac{EUV} spectrometers capable of diagnosing transition region and coronal plasma properties --- including composition --- are required, such as the forthcoming \myac{SOLARC} mission and its spectral imager \myac{EUVST}. These should be complemented by high-cadence \myac{EUV} imaging, for example with \solo/\myac{EUI}/\myac{HRI}, to capture the global coronal dynamics and small-scale structures. On the other side, high-resolution measurements from advanced ground-based telescopes --- such as \myac{GST} or \myac{DKIST} --- are essential for the detection of Alfv\'en waves in the chromosphere as well as for resolving the fine-scale magnetic structures at plume footpoints (below 100\,km) and characterizing the chromospheric environment \citep{alfen_waves_lower_At_1,alfen_waves_lower_At_2,alfen_waves_lower_At_3}. This would provide constraints on the chromospheric wave energy content and polarization, allowing direct comparison with coronal FIP bias measurements and improving the current FIP bias models.



\section*{Acknowledgements}
S.M. acknowledges funding from the Université Paris-Saclay EDOM doctoral scholarship. This work was supported by the Paris Region via the DIM Origines funding. S.M. and M.J. acknowledge support from the European Space Agency (ESA) Archival Research Visitor Programme. 
    D.B. is funded under Solar Orbiter EUI Operations grant number ST/X002012/1 and Hinode Ops Continuation 2022-25 grant number ST/X002063/1.
N.Z.P. is supported by STFC Consolidated Grant ST/W001004/1. 
A.S.H.T. is supported by the ESA Research Fellowship.
\solo is a space mission of international collaboration between ESA and NASA, operated by ESA. We are grateful to the ESA SOC and MOC teams for their support. 
The development of SPICE has been funded by ESA member states and ESA. It was built and is operated by a multinational consortium of research institutes supported by their respective funding agencies: IAS (Centre National d'Etudes Spatiales (CNES), operations lead), STFC RAL (the UK Space Agency (UKSA), hardware lead), GSFC (NASA), MPS (Deutsches Zentrum für Luft- und Raumfahrt (DLR)), PMOD/WRC (Swiss Space Office (SSO)), SwRI (NASA), UiO (Norwegian Space Agency).
The EUI instrument was built by CSL, IAS, MPS, MSSL/UCL, PMOD/WRC, ROB, LCF/IO with funding from the Belgian Federal Science Policy Office (BELSPO/PRODEX PEA 4000112292 and 4000134088), CNES,  UKSA, the Bundesministerium für Wirtschaft und Energie (BMWi) through DLR; and SSO. 
The German contribution to SO/PHI is funded by the BMWi through DLR and by MPG central funds. The Spanish contribution is funded by AEI/MCIN/10.13039/501100011033/ and European Union “NextGenerationEU”/PRTR” (RTI2018-096886-C5,  PID2021-125325OB-C5,  PCI2022-135009-2, PCI2022-135029-2) and ERDF “A way of making Europe”; “Center of Excellence Severo Ochoa” awards to IAA-CSIC (SEV-2017-0709, CEX2021-001131-S); and a Ramón y Cajal fellowship awarded to DOS. The French contribution is funded by CNES.
This work used data provided by the MEDOC data and operations centre (CNES / CNRS / Universite-Paris-Saclay), \url{https://idoc.osups.universite-paris-saclay.fr/medoc/}.
CHIANTI is a collaborative project involving George Mason University, the University of Michigan (USA), University of Cambridge (UK) and NASA Goddard Space Flight Center (USA). 
The authors would like to thank Susanna Parenti, Teodora Mihailescu and Ramada Sukarmadji for fruitful discussions and invaluable insights. 
We also thank Luca Franci and Clara Froment for their coordination of the SOOP observations used in this study and for their valuable input and discussions.
We thank the referees for their constructive comments, which significantly improved the quality and clarity of this work prior to publication.
We acknowledge the use of the following Python librairies: version 7.0 of the SunPy open source software package, 
Astropy, a community-developed core Python package and an ecosystem of tools and resources for astronomy, as well as the Numpy and Scipy libraries.
We acknowledge the use of the following software in this work: 
the \texttt{fiplcr} package for FIP-bias diagnostics using the Linear Combination Ratio method \citep{LCR}; 
EISPAC for Hinode/EIS data analysis \citep{eispac}; 
\texttt{NumPy} \citep{numpy}; 
\texttt{SciPy} \citep{Scipy}; 
\texttt{Matplotlib} \citep{matplotlib}; 
and the \texttt{multiprocess}/\texttt{pathos} libraries for parallel execution \citep{multip}. 




\appendix
\begin{figure*}
    \centering
    \includegraphics[width=1\linewidth]{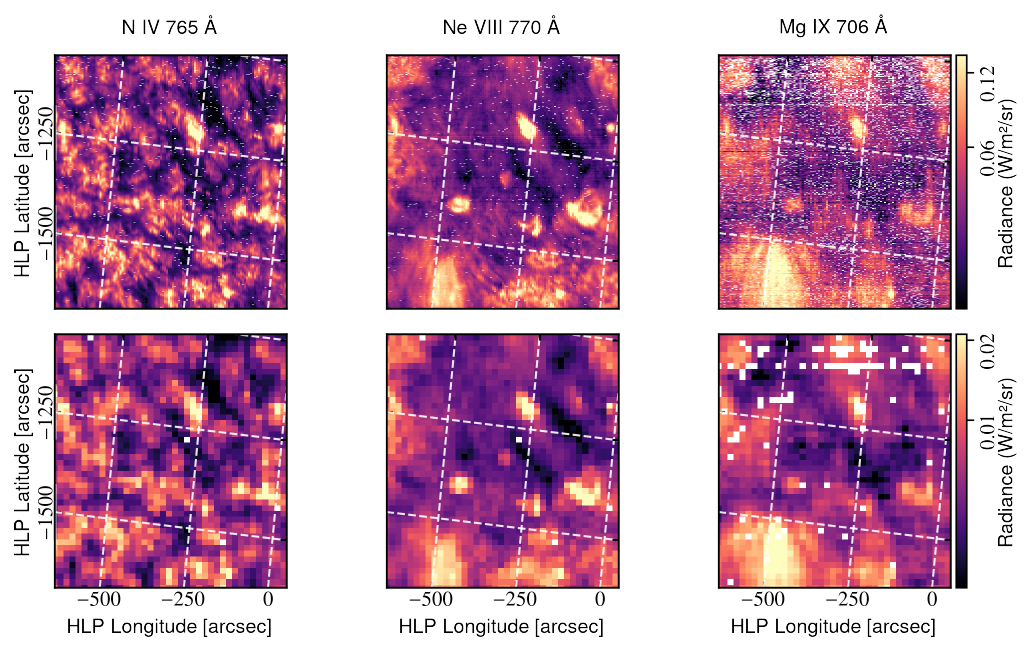}
    \caption{
    \added{Sample \myac{SPICE} radiance maps from \mydateapj[\noyear]{2024,03,31,09,05}, shown for 
    \ion{N}{4}~765.152\,\AA\ (one of the lines used for the composition diagnostics), 
    \ion{Ne}{8}~770.428\,\AA, and \ion{Mg}{9}~706.060\,\AA. 
    The top row displays the full-resolution maps, and the bottom row shows the corresponding binned maps.}
}

    \label{fig:spice_rad_sample}
\end{figure*}

\begin{figure*}[ht]
\centering
\includegraphics[width=1.\linewidth]{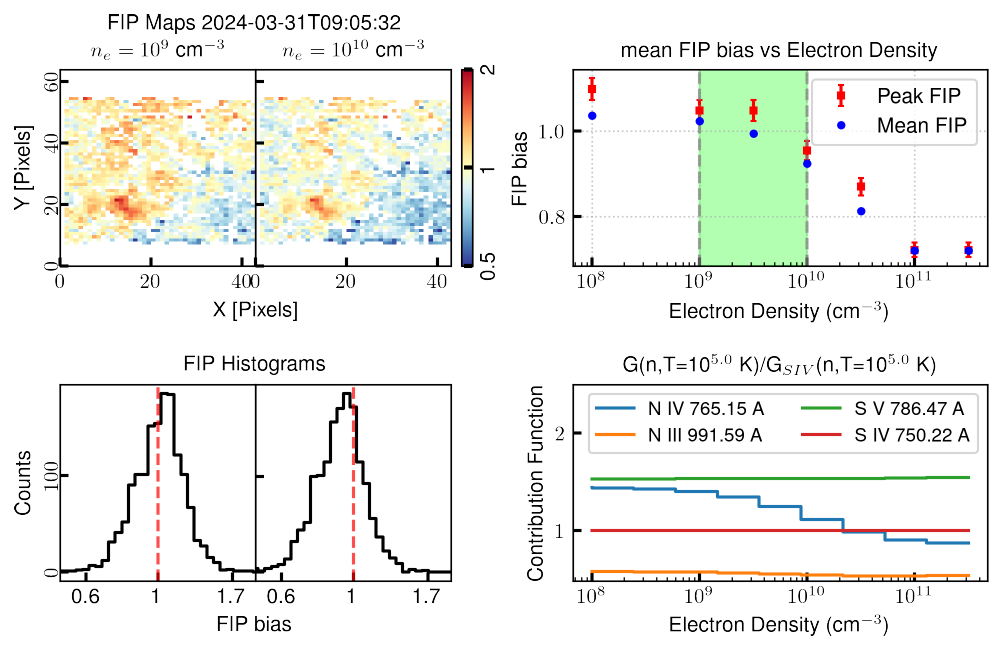}
\caption{
Effect of electron density on sulfur FIP bias derived from \myac{SPICE} data.
\textbf{Left column:} FIP bias maps (top) and corresponding histograms (bottom) for two representative electron densities: $n_e = 10^9\,\mathrm{cm}^{-3}$ and $n_e = 10^{10}\,\mathrm{cm}^{-3}$, highlighting the impact of density assumptions on both spatial distribution and statistical spread of the FIP bias.
\textbf{Top-right:} Mean (blue circles) and histogram peak (red squares) FIP bias values as a function of electron density. Error bars on the peak values denote bin widths. The shaded green area marks the coronal hole density range.
\textbf{Bottom-right:} Normalized contribution functions $G(n, T=10^5\,\mathrm{K})$/$G_{\text{\ion{S}{4}}}(n, T=10^5\,\mathrm{K})$ of selected spectral lines.
}
\label{fig:density_analysis_app}
\end{figure*}

\begin{figure}[ht]
\centering
\includegraphics[width=0.48\linewidth]{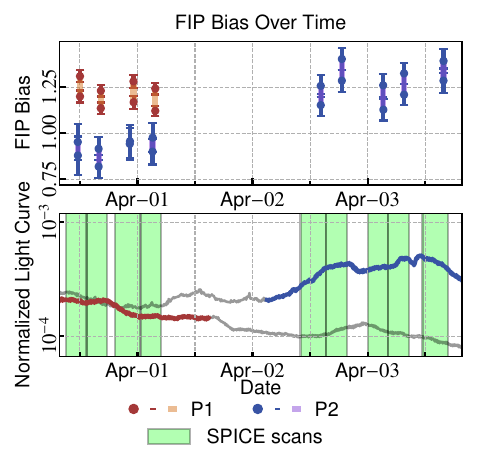}
\caption{
Same as Fig.\,\ref{fig:P1_FIPvsTime}, but now also showing the FIP bias computed assuming an electron density of $10^9\,\mathrm{cm}^{-3}$ (filled markers above each original point). Additional vertical error bars (in orange for ``P1'' and in purple for ``P2'') represent the uncertainty introduced by varying the assumed electron density between $10^9$ and $10^{10}\,\mathrm{cm}^{-3}$, highlighting the impact of density assumptions on the FIP bias.
}
\label{fig:P1_FIPvsTime_den}
\end{figure}
\section{SPICE Binning and Spectral Fitting Procedure}
\label{sec:app__spice_processing}

Binning kernels were adjusted depending on the raster step size. For the 4\arcsec-step rasters, data were binned by 3 pixels in the raster \(x\)-direction (slit scanning direction) and 13 pixels along the \(y\)-direction. For the 6\arcsec-step rasters, only 2 pixels were binned in \(x\)-direction to get a closer resolution between the two study types. This choice maintains a comparable binned pixel size of roughly 12\arcsec\ across both study types and in both spatial directions. \added{Figure~\ref{fig:spice_rad_sample} shows sample \myac{SPICE} radiance maps of \ion{N}{4}, \ion{Ne}{8}, and \ion{Mg}{9}, obtained by fitting the original data (top) and the binned data (bottom).}

Uncertainty estimates for the \myac{SPICE} line intensities were computed using the \texttt{spice\_error()} routine in the \texttt{sospice} Python package, following the methodology detailed in App.~A of \citet{2023A&A...673A..82H}. 

Each spectral window was fitted with a multi-Gaussian function:
\begin{equation}
    F(\lambda; \{I_i\}_{i=1}^{N},\{\lambda_{i}\}_{i=1}^{N}, \{\sigma_i\}_{i=1}^{N},B) = B + \sum_{i=1}^{N} I_i \cdot e^{-\frac{\left(\lambda-\lambda_i\right)^2}{2\sigma_i^2}}
    \label{eq:fit_model}
\end{equation}
where $N$ is the number of emission lines per window, $B$ is the background level, and $I_i$, $\lambda_i$, and $\sigma_i$ represent the intensity, centroid, and width of the $i$-th Gaussian, respectively. Fitting was performed using SciPy’s \texttt{curve\_fit}.

Blended lines crucial for further analysis were treated with constrained fitting. For example, \ion{S}{4}~750.221\,\AA\ is blended with \ion{Mg}{9}~749.552\,\AA. These were disentangled using fixed wavelength offsets relative to other lines of the same ionization state that are not blended, namely \ion{S}{4}~748.393 and \ion{Mg}{9}~706.060. The centroid constraints are described as such:
\begin{align}
    \lambda_{ \text{\scriptsize{\ion{S}{4}~750.221}} } &= \lambda_{ \text{\scriptsize{\ion{S}{4}~748.393}} } + \Delta\lambda_\text{S} \\
    \lambda_{ \text{\scriptsize{\ion{Mg}{9}~749.552}} } &= \lambda_{ \text{\scriptsize{\ion{Mg}{9}~706.060}} } + \Delta\lambda_\text{Mg}
\end{align}
where $\Delta\lambda_\text{S}$ and $\Delta\lambda_\text{Mg}$ are theoretical separations \citep[NIST atomic database version 5.12,][]{NIST_DATABASE,NIST},between the blended and non-blended lines.

Similarly, the weak blended lines \ion{Fe}{3}~985.852, \ion{Ne}{6}~992.683, and \ion{Fe}{3}~994.258 were fitted with centroid constraints relative to \ion{N}{3}~991.577. This approach prevents contamination of the \ion{N}{3} line profile by nearby low-intensity lines and provides a more reliable estimate of the local background. \added{The two \ion{Fe}{3} lines lie at the edges of the spectral window and are therefore not always fully captured.}

\section{Density Effects on the FIP Bias}
\label{sec:app___density_effects}

The FIP bias values computed in this work rely on a fixed choice of electron density. In this section, we detail the assumptions adopted, justify their use in the context of plume fractionation, and examine how varying the density affects the resulting FIP bias.

Since the \myac{SPICE} dataset we used lacked the density-diagnostic lines, we adopted fixed values within the expected \myac{CH} density range. To remain conservative, we selected a relatively high value of $10^{10}\,\mathrm{cm}^{-3}$---representative of the upper limit typically found in the transition region of \myac{CHs} \citep{density_CH_TR,plume_den_TR}. This choice tends to minimize the estimated sulfur FIP bias, as the \ion{N}{4} line used in our diagnostic is density-sensitive and produces lower FIP bias values at higher densities as we are going to show.

With this conservative approach, any fractionation signal observed is a lower bound. If real electron densities are lower, the true FIP bias would be higher. Figure~\ref{fig:density_analysis_app} illustrates this sensitivity. FIP bias maps are shown for the two boundaries of the expected \myac{CH} density range, revealing a consistent trend: lower densities yield higher FIP bias values. This trend is also reflected in the corresponding histograms, where both the peak and mean shift toward higher values as density decreases.

The top-right panel of Fig.\,\ref{fig:density_analysis_app} summarizes this behavior. It shows the mean and peak FIP bias as a function of assumed electron density, across a range from $10^8$ to $10^{11.5}\,\mathrm{cm}^{-3}$. Most values converge near unity within the green-shaded region, which marks the typical \myac{CH} density range. This consistency further supports the plausibility of our fixed-density approach.

Among the selected spectral lines, \ion{N}{4}~765.152\,\AA\ is the density-sensitive line in this collection which is shown in the bottom-right panel of Fig.\,\ref{fig:density_analysis_app}, that shows the normalized theoretical G(n,T) (relative to that of \ion{S}{4}) at $T = 10^5\,\mathrm{K}$. The contribution of the \ion{N}{4} line changes significantly for $n_e > 10^{9.5}\,\mathrm{cm}^{-3}$, while the other lines remain largely unaffected by density. This confirms that the \ion{N}{4} line is the dominant source of density dependence in our FIP diagnostic. For the time being, this method provides the most practical way to identify sulfur fractionation signatures under the current constraints.

To assess how the density assumption affects the temporal evolution of the FIP bias, we take Fig.\,\ref{fig:P1_FIPvsTime}, which displays the time series of the weighted mean FIP bias in ``P1'' and ``P2'', and present in Fig.\,\ref{fig:P1_FIPvsTime_den} the same quantities recalculated using both the upper and lower bounds of the \myac{CH} density range ($10^9$–$10^{10}\,\mathrm{cm}^{-3}$). Under the lower density assumption, FIP bias values in both regions increase, with some reaching above 1.4. In particular, the non-fractionated state of ``P2'' prior to plume appearance is then consistent with \myac{CH} expectations (bias close to unity).

\section{Error estimation} \label{sec:app__errors}
\subsection{Error Estimation from Spectral Line Fitting}
\label{appendix:fit_errors}

Spectral line intensities are derived from Gaussian fitting performed using the \texttt{curve\_fit} function from the \texttt{scipy.optimize} Python library. This routine implements the Levenberg–Marquardt algorithm \citep{levenberg_method,marquardt_algorithm} for non-linear least-squares minimization. Crucially, the fitting procedure incorporates pixel-level statistical uncertainties from the data cube during the fit.

These statistical uncertainties are computed using the \texttt{spice\_error} function from the \texttt{sospice} Python module \citep[see][]{2023A&A...673A..82H}. If spectral binning is applied to improve signal-to-noise, the variances are propagated as:
\begin{equation}
\Delta f_{\text{bin}} = \sqrt{ \frac{1}{N^2} \sum_{i=1}^{N} (\Delta f_i)^2 },
\end{equation}
where \( \Delta f_i \) are the individual pixel uncertainties and \( N \) is the number of pixels combined. This ensures correct weighting during the fit.

The fitting process returns both the best-fit parameters and their 1\(\sigma\) uncertainties derived from the square root of the diagonal of the covariance matrix. For each spectral line, the following are obtained:
\begin{itemize}
  \item \( I_i \) and \( \Delta I_i \): the peak intensity and its uncertainty,
  \item \( \sigma_i \) and \( \Delta \sigma_i \): the Gaussian width (standard deviation) and its uncertainty,
  \item \( \lambda_i \) and \( \Delta \lambda_i \): the line centroid and its uncertainty (not used in radiance calculations).
\end{itemize}

The total radiance \( \mathcal{I}_i \) of the line is computed as:
\begin{equation}
\mathcal{I}_i = \sqrt{2\pi} \cdot I_i \cdot \sigma_i .
\end{equation}

Uncertainties on the radiance are propagated from the fitting parameters via standard first-order error propagation:
\begin{equation}
\frac{\Delta \mathcal{I}_i}{\mathcal{I}_i} = \sqrt{
\left(\frac{\Delta I_i}{I_i}\right)^2 +
\left(\frac{\Delta \sigma_i}{\sigma_i}\right)^2 }.
\end{equation}

These radiance uncertainties \( \Delta \mathcal{I}_i \) are then used as input for the FIP bias uncertainty estimation that are going to be presented in Sec.~\ref{appendix:FIP_errors}.

\subsection{Error Estimation on FIP Bias Values}
\label{appendix:FIP_errors}

Uncertainty in the FIP bias values, as defined in Eq.~\ref{FIP_equation} and based on the combination ratio method \citep{LCR}, is estimated using a first-order approximation. The relative error in the FIP bias $\FIPrepr$ is given by:
\begin{equation}
\frac{\Delta \FIPrepr}{\FIPrepr} =   
\frac{\sum\limits_{i \in (\mathrm{LF})} \alpha_i \Delta \mathcal{I}_i/A_i^P}
           {\sum\limits_{i \in (\mathrm{LF})} \alpha_i \mathcal{I}_i/A_i^P} + 
\frac{\sum\limits_{j \in (\mathrm{HF})} \beta_j \Delta \mathcal{I}_j/A_j^P}
           {\sum\limits_{j \in (\mathrm{HF})} \beta_j \mathcal{I}_j/A_j^P},
\end{equation}
where \( \mathcal{I}_i \) and \( \Delta \mathcal{I}_i \) are the radiance and its uncertainty for line \( i \), \( A_i^P \) are the photospheric abundances, and \( \alpha_i \) and \( \beta_j \) are the coefficients used for the FIP bias calculation (Eq.~\ref{FIP_equation}).

\begin{figure*}[ht]
\centering
\includegraphics[width=1.\linewidth]{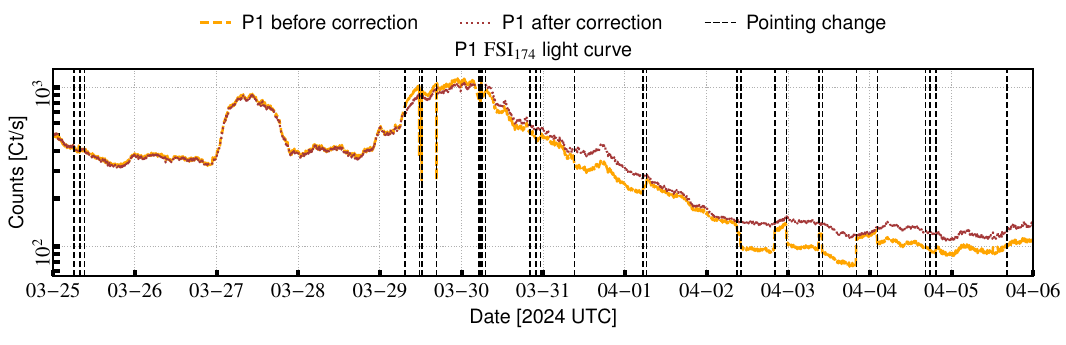}
\caption{Example light curve from region ``P1'' in \FSIFe. Orange: raw intensity. Red: corrected intensity after pointing jump normalization. Dashed vertical lines indicate detected pointing changes.}
\label{fig:light_curve_app}
\end{figure*}

\section{Light Curve Computation Method}
\label{sec:app___light_curves_calculations}

To extract reliable information on temporal variations from \FSIFe\ light curves, despite the absence of absolute radiometric calibration and of time-dependent flat-field correction in EUI data release 6.0, we developed a correction method that normalizes the intensity time series within selected regions of interest (ROIs). This approach compensates for fluctuations caused by spacecraft pointing changes, which shift the ROI across different detector areas with different responses. Our goal is not to retrieve absolute intensities, but to recover consistent, relative intensity variations over time that can be used to track plumes evolution.


\begin{enumerate}
\item For each observation, we compute the mean intensity within the ROI and extract the \myac{WCS} metadata: \texttt{CRVAL1} and \texttt{CRVAL2} (image center coordinates in arcseconds), and \texttt{CDELT1}, \texttt{CDELT2} (pixel scale in arcsec/pixel along $x$ and $y$).

\item We compute the pixel-scale pointing displacement between consecutive images as:
\[
D_{px} = \sqrt{
\left(\frac{\Delta\texttt{CRVAL1}}{\texttt{CDELT1}}\right)^2 +
\left(\frac{\Delta\texttt{CRVAL2}}{\texttt{CDELT2}}\right)^2
}
\]
where $\Delta\texttt{CRVALi}$ is the difference in image center coordinates between two consecutive observations. Thus, $D_{px}$ is the effective pixel displacement between successive pointings. A jump is flagged if $D_{px} > 2$ pixels.

\item Temporal segments between two pointing changes that contain fewer than three measurements are discarded to ensure statistical robustness.

\item The mask is refined iteratively to remove all segments that are too short. Pointing displacements are recomputed using the filtered set until convergence is reached.

\item For each remaining pointing change at index $n$, we compute a correction factor based on the mean intensity before and after the jump using the first light curve (``P1'' in our case):
\begin{equation}
    f_\text{adj} = \frac{I_\text{pre}}{I_\text{post}} = 
    \frac{
        \frac{1}{l} \sum\limits_{k=n-l}^{n-1} I_k^{(0)}
    }{
        \frac{1}{l} \sum\limits_{k=n}^{n+l-1} I_k^{(0)}
    }
\end{equation}
where $I_k^{(0)}$ is the mean intensity of the first contour at time step $k$, and $l=3$ is the averaging window length. This ensures stability in both pre- and post-jump estimates.

\item The factor $f_\text{adj}$ is then applied to all subsequent intensities in all ROIs (e.g. ``P1'' and  ``P2''):
\begin{equation}
    I_m^{(i)} \rightarrow \frac{I_m^{(i)}}{f_\text{adj}}, \quad \forall m \ge n,\; \forall i
\end{equation}
Applying a common correction factor across ROIs is justified if the ROIs are located sufficiently close to each other (within $\sim$500 pixels on the detector), and if large ROI sizes ensure local detector response variations are averaged out.
\end{enumerate}

By applying this correction iteratively across all detected pointing jumps, we obtain smoothed and consistent light curves that more accurately reflect actual coronal variability. This correction is particularly important given the current calibration state of \FSIFe.

Figure~\ref{fig:light_curve_app} shows the light curve from region ``P1'' before (orange) and after (red) correction. Black dashed lines indicate the time steps where a pointing jump was identified and corrected. Several of these discontinuities are significantly reduced by the correction.
\newpage
\bibliography{Bibliography}{}
\bibliographystyle{mnras}
\end{document}